\documentclass[a4paper,11pt]{article}

\usepackage{graphics}
\usepackage{latexsym}
\usepackage{amsmath}
\usepackage{amssymb}
\newcommand{\ap}{\alpha^{\prime}}

\newcommand{\cC}{\mathcal{C}}

\newcommand{\cL}{\mathcal{L}}
\newcommand{\cM}{\mathcal{M}}

\newcommand{\cO}{\mathcal{O}}

\newcommand{\cV}{\mathcal{V}}

\begin{document}
\begin{titlepage}
\thispagestyle{empty}
\begin{flushright}
UT-03-21 \\
hep-th/0306096 \\
June, 2003 
\end{flushright}

\vskip 1.5 cm

\begin{center}
\noindent{\textbf{\Large{Toward Open-Closed String Theoretical Description 
\vspace{0.5cm}\\ of Rolling Tachyon }}} 
\vskip 1.5cm
\noindent{\large{Kazuki Ohmori}\footnote{E-mail: ohmori@hep-th.phys.s.u-tokyo.ac.jp}}\\ 
\vspace{1cm}
\noindent{\small{\textit{Department of Physics, Faculty of Science, University of 
Tokyo}} \\ \vspace{2mm}
\small{\textit{Hongo 7-3-1, Bunkyo-ku, Tokyo 113-0033, Japan}}}
\end{center}
\vspace{1cm}
\begin{abstract}
\baselineskip 5.1mm
We consider how the time-dependent decay process of an unstable D-brane 
should be described in the full (quantum) open-closed string theory. 
It is argued that the system, starting from the unstable D-brane configuration, 
will evolve in time into the time-\textit{in}dependent 
open string tachyon vacuum configuration which we assume to be finite, 
with the total energy conserved. 
As a concrete realization of this idea, we construct a toy model describing 
the open and closed string tachyons which admits such a time-dependent solution. 
The structure of our model has some resemblance to 
that of open-closed string field theory. 
\end{abstract}
\end{titlepage}
\newpage
\baselineskip 6mm


\section{Introduction}\label{sec:introduction}
In a series of papers~\cite{sen12,sen3} Sen has proposed ways to describe 
the classical time evolution of the open string tachyon field $T(t)$ 
on an unstable D-brane, at strictly zero string coupling $g_s=0$. 
Some of the results obtained using the BCFT method~\cite{sen12} were also reproduced from an effective field 
theory~\cite{sen3} and boundary string field theory~\cite{bsft}. 
Since the closed string degrees of freedom, which are responsible for the dissipation of the energy 
liberated from the D-brane, are not taken into account there, 
the energy must be stored in the open string sector in spite of the fact that 
the D-brane is decaying. In order for the energy to be conserved in the settings of~\cite{sen3,bsft}, 
it turns out that the `velocity' of the tachyon field must approach a finite value 
$\dot{T}(t)\to 1$ in the asymptotic future $t\to\infty$. 
This is possible only when the tachyon potential has its minimum at infinity. 
Then, what happens in theories where the minimum of the tachyon potential is at a finite point 
in the field configuration space, such as open string field theory and $p$-adic string theory? 
The authors who have tried to seek rolling solutions in such theories have found the oscillations 
with ever-growing amplitude, typically of the form $T(t)=\sum_na_ne^{nt}$~\cite{MZ,FH,MS}. 
It is possible that these seemingly different behaviors of the tachyon field are related 
to each other through some field redefinition, but at present we do not know precisely 
how to reconcile them. 
Furthermore, it has been suggested that, once the non-zero string coupling $g_s$ is 
turned on, the energy which was initially stored in the open string sector is 
converted to the closed string radiation, no matter how small $g_s$ 
is~\cite{closed,DJ}. 
\medskip

In this paper, we will try to attack the problem of rolling tachyon from the viewpoint of 
open-closed string theory. For simplicity we will restrict ourselves to 
the spatially homogeneous decay of an unstable space-filling D-brane. 
And we assume that the tachyon vacuum is at a \textit{finite} point in the field 
configuration space, in which case much more definite discussion is possible 
than in the case of the runaway potential. 
Then, we expect the following properties for the open string tachyon condensation in 
open-closed string theory: 
\begin{enumerate}
\item
There are two homogeneous and static (\textit{i.e.} constant in spacetime) solutions 
corresponding to the unstable D-brane background and the open string tachyon vacuum where 
the D-brane has disappeared. 
These two configurations should have the same energy density because the energy initially stored 
in the D-brane (=the open string modes) is just transferred to the closed string modes. 
\item
The rolling of the tachyon is described by a time-dependent solution 
which interpolates between the above two vacuum configurations. 
In particular, since the open string sector can lose its energy, 
it would be natural to consider that the tachyon field will eventually relax in 
the (finite) tachyon vacuum. 
\end{enumerate}
As is clear from the appearance of a factor of $\sqrt{\hbar}$ in the open-closed string 
action~(\ref{eq:OCSFTaction}), we must go beyond the classical level to 
incorporate the interactions between open and closed strings.\footnote{For a recent discussion 
about the quantum aspects of Witten's cubic open string field theory, see~\cite{EST}.}  
Since it is quite difficult to deal with the full open-closed string field 
theory (see \cite{9705241,AKT} and references therein), here we consider 
a simple model including tachyons only, 
as a first step toward the ultimate goal. 
In the next section we introduce the model and show that it admits a rolling solution 
satisfying the properties mentioned above. Section~\ref{sec:summary} is devoted to summary and discussion.

\section{A Toy Model for Open and Closed String Tachyons}\label{sec:toymodel}
Let us consider a model described by the following 
action,\footnote{We are working in units where $\ap =1$.}
\begin{eqnarray}
S&=&\int d^Dx \cL=\int d^Dx(\cL_o+\cL_c+\cL_{\mathrm{int}}); \nonumber \\
\cL_o&=&\frac{1}{2}\phi\Box\phi+\frac{1}{2}\phi^2, \nonumber \\
\cL_c&=&\frac{1}{2}\psi\Box\psi+2\psi^2, \nonumber \\
\cL_{\mathrm{int}}&=&-\frac{1}{3}\widetilde{\phi}^3+c_2\widetilde{\phi}\widetilde{\psi}
-\widetilde{\phi}^2\widetilde{\psi}, \label{eq:OCmodelaction}
\end{eqnarray}
where $\phi$ and $\psi$ are the open and closed string tachyon fields respectively, and we have defined 
\begin{equation}
\widetilde{A}(x)\equiv e^{(\log K) \Box}A(x)=K^{\Box}A(x), \label{eq:tilde}
\end{equation}
with $K=2$. 
We take the flat spacetime metric to be $\eta^{\mu\nu}=\mathrm{diag}(-1,1,\cdots ,1)$, 
so that the d'Alembertian is $\Box=-\partial_t^2+\nabla^2$. 
We do not need to specify the spacetime dimensionality $D$. 
$c_2$ is a constant which will be fixed below. 

One may notice that the structure of the action~(\ref{eq:OCmodelaction}), in particular 
the $\widetilde{\phantom{Q}}$-operation~(\ref{eq:tilde}), is reminiscent of the 
open-closed string field theory action~\cite{9705241} 
\begin{eqnarray}
& &\hspace{-4mm}S_{\mathrm{OCSFT}}\sim \left(\langle\Phi|Q_B|\Phi\rangle+g_o\langle\Phi^3\rangle_{\mathrm{disk}}
+g_o^2\langle\Phi^4\rangle_{\mathrm{disk}}\right) +\langle\Psi|(c_0-\bar{c}_0)(Q_B+\bar{Q}_B)|
\Psi\rangle \nonumber \\
& &{}\hspace{1cm}+\sqrt{\hbar}\left(g_c\langle\Psi^3\rangle_{\mathrm{sphere}}+\langle\Psi\rangle_{\mathrm{disk}}
+g_o\langle\Phi\Psi\rangle_{\mathrm{disk}}+g_o^2\langle\Phi^2\Psi\rangle_{\mathrm{disk}}\right)
+\cO(g_o^3,\hbar), \label{eq:OCSFTaction}
\end{eqnarray}
where $\Phi$ and $\Psi$ are open and closed string fields, respectively. 
By truncating the string fields as 
\begin{equation}
|\Phi\rangle=\int\frac{d^Dk}{(2\pi)^D}\phi(k) c_1|k\rangle, \qquad 
|\Psi\rangle=\int\frac{d^Dk}{(2\pi)^D}\psi(k) c_1\bar{c}_1|k\rangle,  \label{eq:expansion}
\end{equation}
and substituting them into the action~(\ref{eq:OCSFTaction}), we would obtain the lagrangian 
whose structure is similar to~(\ref{eq:OCmodelaction}).\footnote{We
have approximately calculated the open-closed transition vertex with the stub and the strip 
omitted, and found that it is roughly estimated as 
$\langle\Phi\Psi\rangle_{\mathrm{disk}}\sim \check{\phi}(x)\check{\psi}(x)$ with 
$\check{A}(x)\simeq\exp((\log 1.352)\Box)A(x)$.
It is quite difficult to compute the open-open-closed vertex 
$\langle\Phi^2\Psi\rangle_{\mathrm{disk}}$ because it includes 
the integration over the moduli space.} However, there should be some discrepancies between 
the resulting action and (\ref{eq:OCmodelaction}). 
In particular, there is no reason why $\psi$, $\psi^3$ and $\phi^4$ terms can be neglected. 
Therefore we consider the action~(\ref{eq:OCmodelaction}) as just a toy model, instead of 
having been derived from the open-closed string field theory action via level truncation. 
\medskip

The vacuum structure of this model can be studied by looking at the potential 
\begin{equation}
\cV=-\cL|_{\phi,\psi =\mathrm{const.}}=-\frac{1}{2}\phi^2-2\psi^2+\frac{1}{3}\phi^3-c_2 \phi\psi
+\phi^2\psi. \label{eq:modelpotential}
\end{equation}
If we eliminate the closed string tachyon field by its equation of motion 
$\psi=\frac{1}{4}(\phi^2-c_2\phi)$, we get the effective potential for $\phi$, 
\begin{equation}
\cV_{\mathrm{eff}}=\frac{1}{8}\phi^2\left\{\left(\phi+\frac{4}{3}-c_2\right)^2
+\frac{8}{3}c_2-\frac{52}{9}\right\}. \label{eq:effpot}
\end{equation}
From this expression, we find that, if we choose $c_2=\frac{13}{6}$, the effective potential 
takes the form $\cV_{\mathrm{eff}}=\frac{\phi^2}{8}\left(\phi-\frac{5}{6}\right)^2$, 
so that $\cV_{\mathrm{eff}}$ has two degenerate vacua at $\phi=0$ and $\frac{5}{6}$. 
With this value of $c_2$, the potential~(\ref{eq:modelpotential}) has three stationary points, 
\begin{equation}
\left.
	\begin{array}{ccll}
\mbox{solution (I)} & : & (\phi^{\mathrm{I}},\psi^{\mathrm{I}})=(0,0), & \cV=0, \\
\mbox{solution (II)} & : & (\phi^{\mathrm{II}},\psi^{\mathrm{II}})=
\left(\frac{5}{6},-\frac{5}{18}\right), & \cV=0, \\
\mbox{solution (III)} & : & (\phi^{\mathrm{III}},\psi^{\mathrm{III}})=
\left(\frac{5}{12},-\frac{35}{192}\right), & \cV=\frac{625}{165888}, 
	\end{array}
\right. \label{eq:vacua}
\end{equation}
where the value of $\cV$ for each solution shows the height of the potential there. 
\medskip

\begin{figure}[htbp]
	\begin{center}
	\scalebox{0.7}[0.7]{\includegraphics{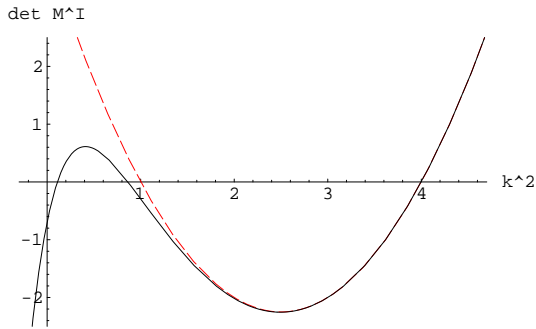}}
	\end{center}
	\caption{$\det \cM^{\mathrm{I}}$ is plotted as a function of $k^2$ (solid line). 
	If the $\phi$-$\psi$ mixing term were absent, the determinant would behave like the 
	dashed line.}
	\label{fig:det1}
\end{figure}

Next we will explain that the solutions (I) and (II) can be regarded as representing 
the D-brane and the open string tachyon vacuum, respectively. 
First, note that there is a $\phi$-$\psi$ mixing term in the action~(\ref{eq:OCmodelaction}) 
at the quadratic level.  
In order to determine the perturbative spectrum around any one of the solutions, we must diagonalize it. 
Let us start with the solution~(I). After the Fourier transformation, 
the quadratic part of the action~(\ref{eq:OCmodelaction}) can be arranged as 
\begin{eqnarray}
& &S_{\mathrm{quad}}^{\mathrm{I}}=-\frac{1}{2}\int\frac{d^Dk}{(2\pi)^D}(\phi(-k),\psi(-k))
\cM^{\mathrm{I}}(k^2)\left(
	\begin{array}{c}
	\phi(k) \\ \psi(k)
	\end{array}
\right), \nonumber \\
& &\cM^{\mathrm{I}}(k^2)=\left(
	\begin{array}{cc}
	k^2-1 & -\frac{13}{6}K^{-2k^2} \\ -\frac{13}{6}K^{-2k^2} & k^2-4 
	\end{array}
\right). \label{eq:massmatrix1}
\end{eqnarray}
The mass spectrum is found by looking for the values of $k^2=-m^2$ 
at which the eigenvalues of the matrix $\cM^{\mathrm{I}}(k^2)$ vanish. This can be equivalently 
accomplished by solving $\det \cM^{\mathrm{I}}(k^2)=0$. Since we cannot solve this equation analytically, 
we resort to the numerical study. We see from Fig.~\ref{fig:det1}, where $\det\cM^{\mathrm{I}}$ is shown 
as a function of $k^2$, that there is a closed string tachyon state with $m_c^2\simeq -4.000$, 
\textit{and} an open string tachyon state with $m_o^2\simeq -0.863$. 
We therefore consider the solution~(I) as the unstable D-brane background. 
We do not concern ourselves about an extra state around $m^2=-0.110$ 
which is not important for our purpose. 

\begin{figure}[htbp]
	\begin{center}
	\scalebox{0.7}[0.7]{\includegraphics{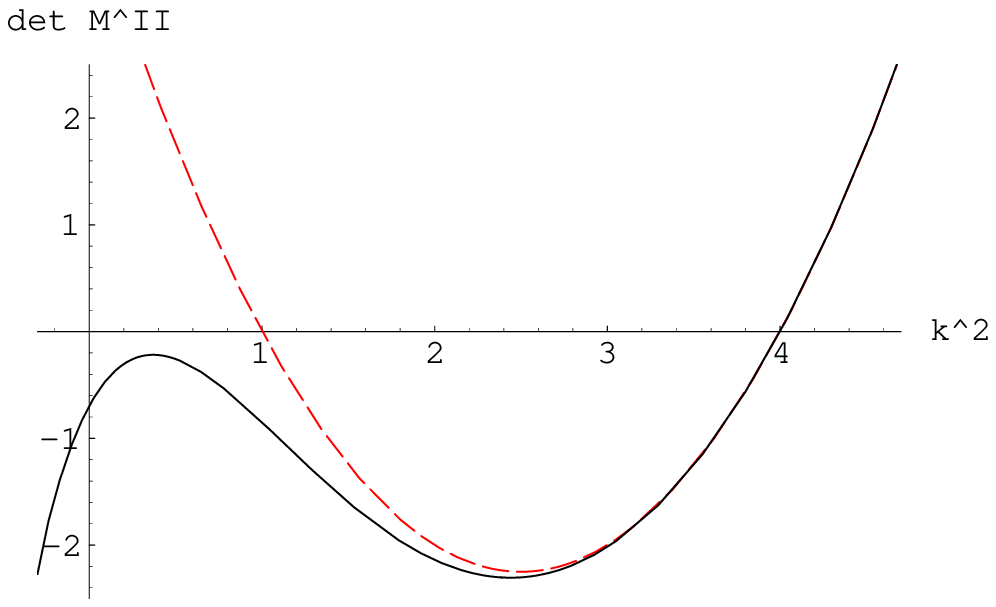}}
	\end{center}
	\caption{$\det \cM^{\mathrm{II}}$ as a function of $k^2$ (solid line).
	There is no solution to $\det\cM^{\mathrm{II}}=0$ around $k^2=1$.}
	\label{fig:det2}
\end{figure}

Now we turn to the solution~(II). When we expand the fields as 
\[\phi=\phi^{\mathrm{II}}+\phi^{\prime},\quad \psi=\psi^{\mathrm{II}}+\psi^{\prime},\] 
the action quadratic in the fluctuation fields $\phi^{\prime},\psi^{\prime}$ becomes 
\begin{eqnarray}
& &S_{\mathrm{quad}}^{\mathrm{II}}=-\frac{1}{2}\int\frac{d^Dk}{(2\pi)^D}(\phi^{\prime}(-k),\psi^{\prime}(-k))
\cM^{\mathrm{II}}(k^2)\left(
	\begin{array}{c}
	\phi^{\prime}(k) \\ \psi^{\prime}(k)
	\end{array}
\right), \nonumber \\
& &\cM^{\mathrm{II}}(k^2)=\left(
	\begin{array}{cc}
	k^2-1+(2\phi^{\mathrm{II}}+2\psi^{\mathrm{II}})K^{-2k^2} & 
	(2\phi^{\mathrm{II}}-\frac{13}{6})K^{-2k^2} \\ 
	(2\phi^{\mathrm{II}}-\frac{13}{6})K^{-2k^2} & k^2-4 
	\end{array}
\right). \label{eq:massmatrix2}
\end{eqnarray}
The determinant of the matrix $\cM^{\mathrm{II}}(k^2)$ is plotted in Fig.~\ref{fig:det2}. 
We see that $\det\cM^{\mathrm{II}}$ vanishes \textit{only once} near $m_c^2\simeq -4.000$. 
This means that the perturbative spectrum around the solution~(II) contains 
the closed string tachyon state, but \textit{not} the open string tachyon. 
Since there is no physical open string state, we consider that 
the solution~(II) corresponds to the background with no D-brane, \textit{i.e.} 
the open string tachyon vacuum. 
\medskip

\begin{figure}[htbp]
	\begin{center}
	\scalebox{1.5}[1.5]{\includegraphics{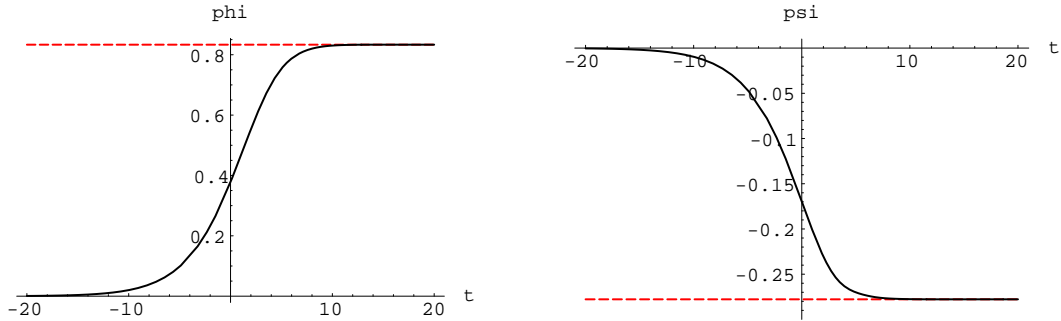}}
	\end{center}
	\caption{The solutions $\widetilde{\phi}(t)$ and $\widetilde{\psi}(t)$ after 7000 steps. 
	The dashed lines show their expectation values 
	$(\phi^{\mathrm{II}},\psi^{\mathrm{II}})$ at the tachyon vacuum. }
	\label{fig:solution}
\end{figure}

Now that we have obtained two vacuum configurations which can be interpreted as 
a D-brane and the tachyon vacuum, we look for a time-dependent solution 
interpolating between them. 
The equations of motion following from the action~(\ref{eq:OCmodelaction}) are 
\begin{eqnarray}
& &(-\partial_t^2+1)K^{2\partial_t^2}\widetilde{\phi}(t)-\widetilde{\phi}(t)^2
+\frac{13}{6}\widetilde{\psi}(t)-2\widetilde{\phi}(t)\widetilde{\psi}(t)=0, \nonumber \\
& &(-\partial_t^2+4)K^{2\partial_t^2}\widetilde{\psi}(t)+\frac{13}{6}\widetilde{\phi}(t)
-\widetilde{\phi}(t)^2=0, \label{eq:modeleom}
\end{eqnarray}
where we have let $\phi$ and $\psi$ be functions only of the time variable $t$. 
The differential operators appearing in~(\ref{eq:modeleom}) can be rewritten as 
the convolution form~\cite{MZ,Volovich}
\begin{eqnarray}
& &(-\partial_t^2+\mu)K^{2\partial_t^2}A(t)=\cC^{(\mu)}[A](t) \nonumber \\
& &\hspace{15mm}\equiv \frac{1}{\sqrt{8 \pi \log K}}
\int_{-\infty}^{\infty}ds\left(-\frac{(t-s)^2}{(4\log K)^2}+\frac{1}{4\log K}+\mu\right)
e^{-\frac{(t-s)^2}{8\log K}}A(s). \label{eq:convolution}
\end{eqnarray}
Then we can numerically solve the `integral equations'~(\ref{eq:modeleom}) 
using the iterative procedure: Given a pair of the $n$-th functions $\phi_n(t)$ and $\psi_n(t)$, 
we can calculate the $(n+1)$-th functions as 
\begin{eqnarray}
& &\widetilde{\phi}_{n+1}(t)=\frac{6}{13}\left(\widetilde{\phi}_n(t)^2-
\cC^{(4)}[\widetilde{\psi}_n](t)\right), \nonumber \\
& &\widetilde{\psi}_{n+1}(t)=\frac{6}{13}\left(-\cC^{(1)}[\widetilde{\phi}_n](t)
+\widetilde{\phi}_n(t)^2+2\widetilde{\phi}_n(t)\widetilde{\psi}_n(t)\right). \label{eq:iteration} 
\end{eqnarray}
If this iterative procedure converges, the set of functions 
$\widetilde{\phi}_{\infty}(t),\widetilde{\psi}_{\infty}(t)$ 
gives a solution to the equations of motion~(\ref{eq:modeleom}). 
Starting with the initial configurations 
\begin{eqnarray}
& &\widetilde{\phi}_0(t)=\left\{\begin{array}{lcc}
	\phi^{\mathrm{I}}=0 & & (t\le 0) \\ \phi^{\mathrm{II}}=\frac{5}{6} & & (t>0)
	\end{array}
\right. , \nonumber \\
& &\widetilde{\psi}_0(t)=\left\{\begin{array}{lcc}
	\psi^{\mathrm{I}}=0 & & (t\le 0) \\ \psi^{\mathrm{II}}=-\frac{5}{18} & & (t>0)
	\end{array}
\right. , \label{eq:initial}
\end{eqnarray}
we have repeated the iterative procedure~(\ref{eq:iteration}) 7000 times, and 
the resulting profiles for $\widetilde{\phi}_{7000}(t)$ and $\widetilde{\psi}_{7000}(t)$ are 
shown in Fig.~\ref{fig:solution}.
We have verified that $\widetilde{\phi}_{7000}(t)$ and $\widetilde{\phi}_{6000}(t)$ (and similarly for 
$\widetilde{\psi}_{7000}(t)$ and $\widetilde{\psi}_{6000}(t)$) completely agree with each other 
at least up to significant six digits for every value of $t$. 
We consider this as strong evidence that $\widetilde{\phi}_n(t)$ and $\widetilde{\psi}_n(t)$ 
really converge in the limit $n\to \infty$. Note that, since $\phi=\widetilde{\phi}$ holds 
within the range where $\phi$ is a constant, the tachyon fields $\phi,\psi$ indeed 
approach their vacuum values with vanishing velocity in this solution. 
\medskip

\begin{figure}[htbp]
	\begin{center}
	\scalebox{1.5}[1.5]{\includegraphics{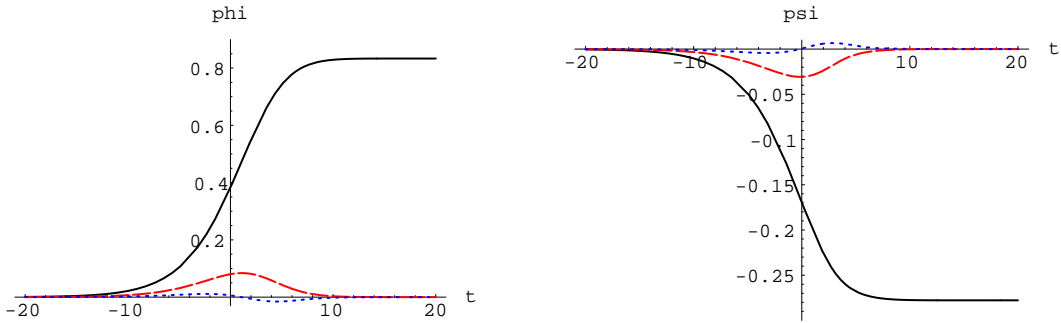}}
	\end{center}
	\caption{Plots of $\phi$ (left), $\psi$ (right) and their derivatives. 
	The solid line shows $\phi(t)$ or $\psi(t)$, the dashed line their first derivatives, 
	and the dotted line their second derivatives.}
	\label{fig:magnitude}
\end{figure}

\begin{figure}[htbp]
	\begin{center}
	\scalebox{1.6}[1.6]{\includegraphics{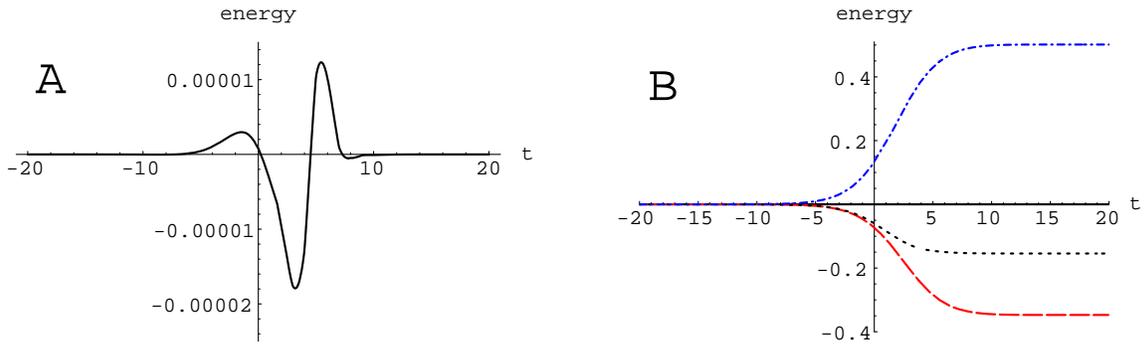}}
	\end{center}
	\caption{A: The total energy $E(t)$ calculated for our numerical solution. 
	B: The energy decomposed into $E_o(t)$ (dashed line), $E_c(t)$ (dotted line) and 
	$E_{\mathrm{int}}(t)$ (dash-dotted line).}
	\label{fig:energy}
\end{figure}

We will now check whether the energy is conserved in time for the solution obtained above. 
The energy $E(t)$ is calculated by the following formula~\cite{MZ}
\begin{eqnarray}
& &E(t)=E_o(t)+E_c(t)+E_{\mathrm{int}}(t); \nonumber \\
& &E_o(t)=-\cL_o+\sum_{l=1}^{\infty}\sum_{m=0}^{2l-1}(-1)^m\left(\frac{\partial\cL}
{\partial\phi_{2l}}\right)_m\phi_{2l-m}, \nonumber \\
& &E_c(t)=-\cL_c+\sum_{l=1}^{\infty}\sum_{m=0}^{2l-1}(-1)^m\left(\frac{\partial\cL}
{\partial\psi_{2l}}\right)_m\psi_{2l-m}, \nonumber \\
& &E_{\mathrm{int}}(t)=-\cL_{\mathrm{int}}, \label{eq:energy}
\end{eqnarray}
where the subscripts denote the number of time-derivatives, 
$A_n(t)\equiv\frac{\partial^n}{\partial t^n}A(t)$. 
Although this formula contains infinite sums, we can obtain a good approximation to it by ignoring 
higher derivative terms, because they have negligibly small orders of magnitude as compared 
to $\phi$ and $\psi$ themselves,\footnote{This also shows that the profile of 
$\widetilde{\phi}$ ($\widetilde{\psi}$) is almost 
the same as that of $\phi$ ($\psi$).} as shown in Fig.~\ref{fig:magnitude}. 
Here we keep only up to the fourth derivatives. 
Note that it is meaningful to 
compare $\phi$ (or $\psi$) with its derivatives since the time variable $t$ has 
been made dimensionless by setting $\ap =1$. 
The total energy $E(t)$ calculated this way is illustrated in Fig.~\ref{fig:energy}A. 
If we compare it with Fig.~\ref{fig:energy}B where the constituents $E_o,E_c,E_{\mathrm{int}}$ 
defined in~(\ref{eq:energy}) are shown separately, it is clear that highly non-trivial 
cancellations do occur among $E_o,E_c$ and $E_{\mathrm{int}}$. 
We thus conclude that the total energy is constant ($=0$) in time. 
However, as pointed out in~\cite{MS}, we do not know how to separate the total energy 
into the open and the closed string sectors. 
It would be interesting if we can see the flow of the energy from open to closed strings. 
\medskip

\begin{figure}[htbp]
	\begin{center}
	\scalebox{1.5}[1.5]{\includegraphics{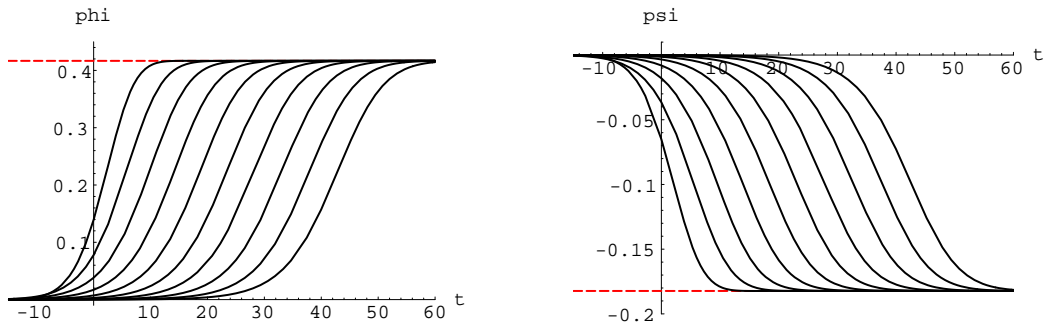}}
	\end{center}
	\caption{$\widetilde{\phi}_n(t)$ (left) and $\widetilde{\psi}_n(t)$ (right) for 
	$n=10,20,30,\ldots ,100$ from left to right. The dashed lines show their expectation values 
	$(\phi^{\mathrm{III}},\psi^{\mathrm{III}})$ at the solution~(III).}
	\label{fig:flee}
\end{figure}

We have also investigated whether we can construct a time-dependent solution 
which interpolates between the solutions~(I) and (III) using the iterative procedure, starting with 
\begin{eqnarray}
& &\widetilde{\phi}_0(t)=\left\{\begin{array}{ccc}
	0 & & (t\le 0) \\ \phi^{\mathrm{III}}=\frac{5}{12} & & (t>0)
	\end{array}
\right. , \nonumber \\
& &\widetilde{\psi}_0(t)=\left\{\begin{array}{ccc}
	0 & & (t\le 0) \\ \psi^{\mathrm{III}}=-\frac{35}{192} & & (t>0)
	\end{array}
\right. . \label{eq:initial2}
\end{eqnarray}
From the result shown in Fig.~\ref{fig:flee},  
we see that the tachyon profiles tend to $\phi(t)\equiv\psi(t)\equiv 0$ 
as $n$ becomes large. Such a behavior is natural because there should be no energy-conserving 
solution connecting these two vacua in time.

\section{Summary and Discussion}\label{sec:summary}
In this paper we have argued that in the full open-closed string theory, 
the spatially homogeneous decay process of a 
space-filling unstable D-brane should be described by a time-dependent solution 
which interpolates between two static and energetically degenerate field configurations 
representing the unstable D-brane and the open string tachyon vacuum. 
In particular, to realize the situation in which the tachyon field converges to some finite 
value, it was essential to incorporate the bulk closed 
string modes because the open string sector must dissipate the energy which amounts to 
the D-brane tension through the couplings to closed strings. 
We have constructed and investigated a simple model describing the open and closed string 
tachyons, and found that (i)~this model has two degenerate vacua which can be interpreted 
as the unstable D-brane and the tachyon vacuum, (ii)~there exists a time-dependent 
solution which interpolates in time between them, and (iii)~the total energy 
is conserved during the decay process described by this time-dependent solution. 
Hence it seems that some features expected of the rolling tachyon 
in open-closed string theory are indeed captured by our model. 
\medskip

There are many things to be studied further. We list some of them below. 
\begin{itemize}
\item
Among others, our analysis based on the toy model~(\ref{eq:OCmodelaction}) is far from complete. 
In particular, there is no plausible justification for neglecting various higher-order interactions 
and for truncating the string fields to the tachyons. 
It would be interesting if we find that the time-dependent solution obtained in 
section~\ref{sec:toymodel} persists to exist 
after reinforcing our model by including higher-order interactions and 
other massless or massive fields. 
\item
We have focused on the spatially homogeneous decay of 
a space-filling D-brane. In the case of the inhomogeneous decay 
or the decay of a lower-dimensional D-brane, 
the closed strings emitted from the decaying D-brane will be propagating 
outward in the transverse space. In this sense, the asymptotic state of the 
closed string field is not static. 
Nevertheless, we do expect that the open string field reaches 
the tachyon vacuum configuration sometime and ceases to evolve in time thereafter. 
\item
We have identified the solution~(II) with the open string tachyon vacuum 
by looking at the fluctuation spectrum around it. 
It is desirable to see that the energy stored in the open string field 
is lower at the solution~(II) than at the solution~(I). 
As mentioned in section~\ref{sec:toymodel}, however, it is not known how to 
split the total energy into open and closed string sectors in our tachyon model. 
It may become possible if we can incorporate the metric degrees of freedom in our model, 
as in~\cite{DJ}. 
\item
It seems that the convolution form of the equations of motion is difficult to relate to 
the `initial value problem'~\cite{sen12,MZ}. 
Our numerical solution corresponds to the rolling from the top of the potential 
with vanishing initial velocity, and has no parameter to be chosen freely. 
In fact, we have encountered a similar situation in seeking for static lump or kink solutions 
in (super)string field theory~\cite{lk}, where we did not treat it as a 
boundary value problem, and the width of the lump (kink) seemed to be determined automatically. 
\end{itemize}
We hope that our investigation presented here sheds new light on the problem of 
the time-dependent decay of unstable D-brane systems. 


\section*{Acknowledgments}
I am grateful to T.~Eguchi, I.~Kishimoto, N.~Moeller, M.~Schnabl and E.~Watnabe for 
valuable discussions. 
I would also like to thank M.~Fujii, Y.~Matsuo, R.~Nobuyama, K.~Sakai, Y.~Tachikawa 
and H.~Takayanagi for useful conversations. 
The numerical computations in this work were performed using \textit{Mathematica}. 
This work is supported by JSPS Research Fellowships for Young Scientists. 



\end{document}